\documentstyle[epsfig]{elsart}
\newcommand{\vsig}{\mbox{\boldmath $\sigma$ \unboldmath}}
\newcommand{\veps}{\mbox{\boldmath $\epsilon$ \unboldmath}}
\newcommand{\valf}{\mbox{\boldmath $\alpha$ \unboldmath}}
\newcommand{\vtau}{\mbox{\boldmath $\tau$ \unboldmath}}
\newcommand{\vpi}{\mbox{\boldmath $\pi$ \unboldmath}}

\newcommand{\vgamma}{\mbox{\boldmath $\gamma$ \unboldmath}}
\begin{document}

\begin{frontmatter}
\title{Nucleonic resonance effects in the $\phi$ meson photoproduction}
\author[ORSAY]{Qiang Zhao},
\author[ORSAY]{J.-P. Didelez},
\author[ORSAY]{M. Guidal},
\author[SACLAY]{B. Saghai} 
\address[ORSAY]{Institut de Physique Nucl\'eaire, F-91406 Orsay 
Cedex, France}
\address[SACLAY]{Service de Physique Nucl\'eaire, DAPNIA-DSM, 
CEA/Saclay,\\
F-91191 Gif-sur-Yvette, France}
\begin{abstract}
The process $\gamma p \to \phi p$  close to threshold is investigated 
focusing on the role played by the {\it s}- and {\it u}-channel 
nucleonic resonances.
For this purpose, a recent quark model approach, based on the 
$SU(6)\otimes O(3)$ symmetry with an effective Lagrangian, is extended
to the $\phi$ meson photoproduction.
Another non-diffractive process, the {\it t}-channel $\pi^0$ exchange, 
is also included.
The diffractive contribution is produced by the {\it t}-channel Pomeron 
exchange.
Contributions from non-diffractive {\it s}- and {\it u}-channel 
process are found small in the case of cross sections 
and polarization observables at forward angles. 
However, backward angle polarization asymmetries show high sensitivity
to this non-diffractive process. 
Different prescriptions to keep gauge invariance for the Pomeron exchange amplitudes are investigated. 
Possible deviations from the exact $SU(6)\otimes O(3)$ symmetry, 
due to the configuration 
mixing, are also discussed. 
\end{abstract}
\end{frontmatter}


\newpage
%  %%  %%  %%  %%  %%  %%  %%  INTRODUCTION
%
\section{ Introduction}

A well established feature in vector meson photoproduction
at low momentum transfer and high energies is that the diffractive
scattering governs the reaction mechanism~\cite{Bauer}. 
With the advent of the new high intensity beam
facilities, like JLAB, ELSA, GRAAL, SPring-8, the study of this field in 
both low energy ($E_{\gamma} \sim 2$ GeV) and/or
high momentum transfer ($-t \geq 1$ (GeV/c)$^2$) becomes possible. 
In these regimes, deviations
from pure diffractive phenomena are expected, which should show up, 
especially in polarization observables.

In recent works, the diffractive component is dealt with as a
{\it t}-channel Pomeron exchange~\cite{Donnachie,Laget,Haakman,lee}, where
the Pomeron is treated as a $C=+1$ isoscalar photon.
For the non-diffractive contributions, various sources have been 
explored. In Ref.~\cite{Williams},
the {\it t}-channel pseudoscalar mesons ($J^\pi = 0^-$, i.e. $\pi^0$ and 
$\eta$) exchanges are in general found 
to produce small contributions at small 
momentum transfers. Here, it will be interesting to study the 
nature of the parity exchange at low energy, namely, the interplay
between diffractive scattering which
is intrinsically a {\it natural} parity exchange and the pseudoscalar 
meson exchange which is an {\it unnatural} parity exchange process.

Another exciting topic concerns the possible  
violation mechanisms of the Okubo-Zweig-Iizuka (OZI) rule~\cite{OZI}.
Such a violation has recently been reported in the $\overline{p}n$ and
$p\overline{p}$ annihilation experiments~\cite{ozi-vio}.
Theoretically, these phenomena can not be explained by 
conventional approaches based on the two-step OZI allowed
final state interaction with the intermediate kaon formation.
Various models have been proposed to account for this, 
initiating the search for strangeness
components in nucleons {\it via} the $\phi$ meson photoproduction near 
threshold~\cite{Henley,Ellis,Titov}.
Then, one needs to investigate on the one hand OZI evading processes
and on the other hand the role of the strangeness content of the proton
{\it via} $s\overline{s}$ knock-out reactions.

Another important component in the reaction mechanism at low energies
is expected to be the contributions from {\it s}- and {\it u}-channel
resonance excitations. 
At present time, there is no systematic investigation of the
role played by the resonances in the $\phi$ meson photoproduction,
which, however, might be 
another important non-diffractive source contributing near threshold.
Although a recent work~\cite{lee98} taking into account
the {\it s}- and {\it u}-channel $\phi NN$ coupling 
has been carried out, the contributions from 
resonances have not been included there. In fact, at hadronic level, 
the unknown $\phi NN^*$ couplings have been the barrier
to go further to include the resonances since one has to introduce 
at least one parameter for each $\phi NN^*$ coupling vertex, 
therefore, a large number of parameters will appear in the theory. 
On this point, 
the quark model approach shows great advantages: in the 
exact $SU(6)\otimes O(3)$ symmetry
limit, the quark-vector-meson interactions can be described by the 
effective Lagrangian with only two parameters.
The main purpose of the present work is, therefore, to
focus on this aspect using a quark-model based effective Lagrangian 
approach in line with a recent work~\cite{plb98,prc98}. 
 Here, we wish to emphasize that in the latter work, 
a very preliminary study of the resonance 
contributions to the $\phi$ meson photoproduction was performed
and, especially, the dominant diffractive contribution was not included. 
Therefore, it was impossible to constrain the parameters there and 
an approximate estimation was made for only the differential cross
section from the resonance contributions. 
In this work, 
the dominant diffractive process has been introduced 
consistently with the quark model approach, and a search is performed
to constraint the parameters. This set of ingredients is expected
to make possible a reliable estimation of the small, but significant,
contributions from the nucleonic resonances to the reaction mechanism.
Our results show that the resonances (with mass $M_R \le$ 2.0 GeV) play
a non-negligible role, especially in polarization observables, in spite of
the rather high $\phi$ meson production threshold 
($E^{thres}_{\gamma} \sim 1.57$ GeV corresponding to the total c.m. energy 
$\sqrt{s}\sim 1.96$ GeV).

As a first step of study in the energy region near threshold, 
we employ a non-relativistic quark potential to describe the 
quark motion.  As it has been shown in Ref.~\cite{RCQM},
the non-relativistic formulation remains a viable approximation
due mainly to the effective parameters such as the constituent quark 
mass. To partly cure the shortcoming arising from the non-relativistic 
quark potential, a widely used Lorentz boost is introduced in the 
spatial integrals~\cite{zpli95}. 

For the pure diffractive process, our formalism embodies 
a model treating the Pomeron like a $C=+1$ isoscalar 
photon~\cite{Donnachie,lee}.
It is the extrapolation from the high
energy region ($E_\gamma = 6.45$ GeV) down to the low energy region. 
Thus, no free parameters are introduced. 
The resonance contributions drop quickly with the energy increasing, 
and almost vanish when the photon energy goes 
above roughly 2.8 GeV. 
So, at higher energies, the $\phi$ meson photoproduction cross section 
is generated by the diffractive process. 
Therefore, with such a Pomeron exchange term, this model can be applied 
to the $\phi$ meson production from threshold up to the 
higher energy region (i.e. $E_\gamma\sim 10$ GeV) 
where the {\it s}- and {\it u}-channel contributions
are negligible. 
Although our interest is to study the energy region near threshold, 
we emphasize that 
a reliable estimation of the diffractive contribution 
in this energy region must be 
the prerequisite to further investigations, especially, when 
focusing on the small non-diffractive contributions.

The contribution from $\pi^0$ exchange is also included in this 
work. However, we find that at $E_\gamma=2.0$ GeV, 
it is quite negligible in 
comparison with not only the Pomeron exchange in the small $|t|$ region, 
but also the resonance contributions in the large $|t|$ region.
We do not include the amplitude for the $\eta$ exchange in this work,  
since although the $\phi$ meson has 
larger decay branching ratio for $\phi\to \eta\gamma$ than 
for $\phi\to \pi^0\gamma$ which means that the $\phi\eta\gamma$ 
coupling is stronger than the $\phi\pi^0\gamma$ coupling, 
a recent analysis~\cite{LS-1} of the $\eta$ photoproduction
shows that  $g_{\eta NN}$ is smaller than $g_{\pi NN}$ by roughly
a factor of 7, 
which leads to negligible effects due to the $\eta$ exchange  
in the $\phi$ production. 

In section 2, we present  
the effective Lagrangian for the quark-vector-meson coupling and 
the amplitude for the Pomeron and $\pi^0$ exchanges. In section 3, the 
numerical results for the differential cross section and  
polarization observables are reported. 
As a test of the model, we present the total cross section up 
to $E_\gamma=10$ GeV in comparison with the experimental data.
We also present comparisons, for the three density matrix elements
$\rho^0_{00}$, $Re\rho^0_{10}$ and $\rho^0_{1-1}$, between our 
predictions and the experimental 
results, showing that the diffractive contribution has been 
treated in a reliable manner.
The predictions for the three density matrix elements are presented 
at $E_\gamma=2.0$ GeV. Possible OZI suppression effects 
are also discussed.
In section 4, Pomeron exchange amplitudes, 
due to different schemes of taking into account the
gauge invariance, are investigated. 
Conclusions are given in section 5.
%
%%%%%%%%%%%%%%%%%%%%%%%%%%%
%
%
\newpage
%  
%
%  %%  %%  %%  %%  %%  %%  %% FORMALISM
%
\section{ Formalism }
In Ref.~\cite{prc98}, the quark model approach to 
 vector meson photoproduction with an effective Lagrangian 
in the resonance region has been developed. In the following Subsection
we summarize the main points of that approach and extend it to the
$\phi$ meson photoproduction process.
Notice that in Ref.~\cite{prc98} the latter process was investigated
only qualitatively, for the following reasons: 
i) the diffractive Pomeron exchange,
which plays a dominant role in the reaction mechanism,
is absent in that work, ii) this latter shortcoming prevents any
parameter search, iii) also, the polarization observables can not
be investigated within Ref.~\cite{prc98} formalism.  

In the remaining two subsections, 
the diffractive Pomeron exchange and the non-diffractive $\pi^0$
exchange are presented.
%
%  %%  %%  %%  %%  %%  %%  %% EFFECTIVE LAGRANGIAN
%
\subsection{ Effective Lagrangian for quark-vector-meson coupling}
The formalism used here is based
on the $SU(6)\otimes O(3)$ symmetry for the 3-quark 
baryon system. We suppose that the $\phi$ meson is produced 
from an elementary process described by an effective Lagrangian 
where the $\phi$ meson is treated as a point-like elementary 
particle.
As we will discuss later, deviations from this symmetry are a valuable
source of information about the structure of the intervening hadrons.

The effective Lagrangian introduced for the quark-meson coupling has
the following form:
\begin{eqnarray} \label{3.0}  
L_{eff}=-\overline{\psi}\gamma_\mu p^\mu\psi+\overline{\psi}\gamma_
\mu e_qA^\mu\psi +\overline{\psi}(a\gamma_\mu +  
\frac{ib\sigma_{\mu\nu}q^\nu}{2m_q}) \phi^\mu_m \psi,
\end{eqnarray} 
where $\psi$ and $\overline{\psi}$ represent the quark and anti-quark
fields, respectively, and $\phi^\mu_m $ denotes the vector meson 
field. The two parameters, 
$a$ and $b$ represent the vector and tensor couplings 
of the quark to the vector meson, respectively,
and $m_q=330$ MeV is the constituent quark mass.

With the above effective Lagrangian, at tree level, the non-diffractive
transition amplitudes can be expressed as the sum of the 
contributions from {\it s}-, {\it u}-, and 
{\it t}-channel: 
\begin{eqnarray}  
M_{fi}=M^s_{fi}+M^u_{fi}+M^t_{fi} \ .  
\label{3.1}  
\end{eqnarray} 

Given that the {\it t}-channel contribution, $M^t_{fi}$, is proportional 
to the charge of the final state meson, it does not contribute 
in the case of neutral $\phi$ meson photoproduction. 
The transition amplitudes from {\it s}- and 
{\it u}-channel
can be written as:
\begin{eqnarray}  
M^{s+u}_{fi}&=&i\omega_\gamma\sum_{j}\langle N_f|H_m|N_j\rangle\langle   
N_j|\frac{1}{E_i+\omega_\gamma-E_j}h_e|N_i\rangle\nonumber\\  
&&+i\omega_\gamma\sum_{j} \langle N_f|h_e\frac{1}{E_i-\omega_\phi-E_j}  
|N_j\rangle\langle N_j|H_m|N_i\rangle, 
\label{3.2}  
\end{eqnarray} 
with 
$H_m=-\overline{\psi}(a\gamma_\mu +\frac{ib\sigma_{\mu\nu}  
q^\nu}{2m_q}) \phi^\mu_m \psi$ for the quark-meson coupling vertex, and
\begin{eqnarray}  
h_e=\sum_{l}e_l{{\bf r}_l\cdot{\veps \hskip -0.16 cm }_\gamma}(1-\valf\cdot  
{\hat{\bf k}})e^{i{\bf k\cdot r}_l},~
{\bf{\hat k}}=\frac{\bf k}{\omega_\gamma}.
\end{eqnarray} 
Here, $l=1,2,3$, denotes the three quarks of the initial or final 
state nucleons, and $e_l$ is the charge of the $l$th quark. 
The matrix, 
$\valf=\gamma^0 \vgamma$,
 where $\gamma^0$ and $\vgamma$ are the Dirac
matrices. 
The kinematic variables are, 
${\bf k}$: the momentum of the incident photon; 
$\omega_\gamma$: the photon energy; 
$\omega_\phi $: the energy of the outgoing meson; 
$E_j$: the energy of the intermediate state in the {\it s}- and 
{\it u}-channels; $E_i$: the energy of the initial state nucleon.
It should be noted that in Eq.~(\ref{3.2}), we have omitted 
the contact term derived from the {\it s}- and {\it u}-channels since 
it is proportional to the charge of the outgoing meson,
therefore, vanishes in the neutral meson production processes.

The explicit expressions for the longitudinal and transverse
{\it s}- and {\it u}-channel transition matrices have been 
derived in Ref.~\cite{prc98}. 
%%
%%%%%%%%%%%%%%%%%%%%%%%%%%%%%% TABLE I
%%
\begin{table}[hb]
\caption{ Resonances in the {\it s}-channel, with their 
assignments in the $SU(6)\otimes O(3)$ symmetry limit, are given in
the first and second columns, respectively. The masses ($M_R$) and 
and total widths ($\Gamma_T$), used in this work, given in columns
third and fourth, respectively, are taken from Ref.~\cite{PDG}.
} 
\protect\label{tab:reso}
\begin{center}
\begin{tabular}{cccc}
\hline
Resonances & $SU(6)\otimes O(3)$ & $M_R$ (MeV) & $\Gamma_T$ (MeV) \\
\hline
$P_{11}(1440)$ & $N(^2S^\prime_S)_{{\frac 12}^+}$ & 1440 & 350\\
$D_{13}(1520)$ & $N(^2P_M)_{{\frac 32}^-}$ & 1520 & 120\\
$S_{11}(1535)$ & $N(^2P_M)_{{\frac 12}^-}$ & 1535 & 150\\
$F_{15}(1680)$ & $N(^2D_S)_{{\frac 52}^+}$ & 1680 & 130\\
$P_{11}(1710)$ & $N(^2S_M)_{{\frac 12}^+}$ & 1710 & 100\\
$P_{13}(1720)$ & $N(^2D_S)_{{\frac 32}^+}$ & 1720 & 150\\
$P_{13}(1900)$ & $N(^2D_M)_{{\frac 32}^+}$ & 1900 & 400\\
$F_{15}(2000)$ & $N(^2D_M)_{{\frac 52}^+}$ & 2000 & 450\\
\hline
\end{tabular}
\end{center}
\end{table}
The transition amplitudes for each resonance in the {\it s}-channel
below 2 GeV are included explicitly, 
while the resonances above 2 GeV with a given quantum number $n> 2$ 
in the harmonic oscillator basis of the quark model
are treated as degenerate. 
The contributions from   
the {\it u}-channel resonances are divided into two parts as well. The
first part contains the baryons with   
the quantum number $n=0$, which includes the spin 1/2 states, 
and the spin 3/2 resonances. Since   
the mass splitting between the spin 1/2 and spin 3/2 resonances   
with $n=0$ is significant, they have to be treated separately. 
 The second part in the {\it u}-channel comes from the excited   
resonances with quantum number $n\ge 1$. As the contributions  
 from the {\it u}-channel resonances are not sensitive to the 
precise mass positions, they can also be treated as degenerate. 
In the $\phi$ meson photoproduction,  
because of the isospin conservation,
the $\phi$ meson photoproduction gets contributions only
from isospin 1/2 resonances. 
Therefore, only the nucleon pole term  ($n=0$)
and those intermediate excited states ($n>0$) with isospin 1/2 contribute
in this reaction. Also, the Moorhouse selection rule~\cite{moorhouse}
suppresses 
those states belonging to representation ${\bf (70, ^48)}$ 
from contributing in the photon excitations of 
the proton target. Therefore, in the NRCQM 
symmetry limit, there are only 8 intermediate nucleonic resonances 
appearing in the {\it s}-channel with $n\le 2$. In Table~\ref{tab:reso}, 
the NRCQM wave-functions of these resonances are presented with their 
masses $M_R$ and total widths $\Gamma_T$.

The transition amplitudes can then be expressed in terms of the 12 
independent helicity amplitudes, which are related to the 
spin observables and the density matrix 
elements~\cite{tabakin,schilling}.
 
The general transition amplitude for the {\it s}-channel
excited states in the helicity space has the following form: 
\begin{eqnarray}  
H^J_{a\lambda_V}=\frac{2M_R}
{s-M_R(M_R-i\Gamma({\bf q}))}   
h^J_{a\lambda_V},    
\label{6.1}    
\end{eqnarray}  
where $\sqrt{s}=E_i+\omega_\gamma=E_f+\omega_m$ is the total energy   
of the system, $E_i$ and $E_f$ are the energies of the nucleons in the 
initial and final states, respectively, $h^J_{a\lambda_V}$ are 
the helicity amplitudes, and $\Gamma({\bf q})$, 
which is a function of the final state 
momentum ${\bf q}$, 
denotes the momentum dependence of the 
total width of the resonance $\Gamma_T$~\cite{zpli95}. 

The differential cross section has the expression:
\begin{eqnarray}
\frac{\mbox{d}\sigma}{\mbox{d}t}=
\frac{\alpha_e (E_f+M_N)(E_i+M_N)}{16 s|{\bf k}|^2}\frac 12
\sum^4_{a=1}\sum_{\lambda_V=0,\pm 1} |H_{a\lambda_V}|^2,
\end{eqnarray}
where $M_N$ represents the mass of the nucleon, and ${\bf k}$ denotes 
the momentum of the incoming photon in the c.m. system.

In the harmonic oscillator basis, 
the factor $e^{-\frac{{\bf q}^2+{\bf k}^2}{6\alpha^2}}$ 
comes from  
the integrations over the 3-quark baryon wave-functions, 
and plays a role  
like a form factor for the quark-meson 
and quark-photon vertices~\cite{zpli93_compton}.
Here, $\alpha$=410 MeV 
is the commonly used value for the harmonic oscillator strength, and
no additional ``cut-off'' parameter is needed. 
Thus, only two parameters, $a$ and 
$b$ in Eq.~(\ref{3.0}) are introduced by the effective 
Lagrangian for the 
{\it s}- and {\it u}-channel non-diffractive $\phi$ production. 
It should be noted that the gauge invariance of the amplitudes 
from the effective Lagrangian has been fixed. 
Moreover, the above factor is implemented with the Lorentz boost
to take into account the relativistic effects.

%
%  %%  %%  %%  %%  %%  %%  %% DIFFRACTIVE POMERON ...
%
\subsection{{\it t}-channel diffractive Pomeron exchange }
We use the Pomeron exchange model by Donnachie and 
Landshoff~\cite{Donnachie} to produce the diffractive 
contribution in this work. 
In the model, the Pomeron mediates the long range interaction 
between a confined quark and a nucleon. 
Although the nature of the Pomeron 
exchange is still unclear, it has been shown that the Pomeron 
exchange based on the Regge phenomenology is one of the 
most successful approaches to high energy elastic scattering. 
Also, it has been shown that the Pomeron behaves 
rather like a  $C=+1$ isoscalar photon. 

With the Pomeron-photon analogy picture,
the Pomeron-nucleon coupling is described by the 
vertex: 
\begin{eqnarray}
F_{\mu}(t)= 3\beta_0\gamma_{\mu}f(t),
\end{eqnarray}
where $-t$ is the Pomeron momentum squared, 
$\beta_0$ gives the strength of the coupling of the single 
Pomeron to a light constituent quark. 
$f(t)$  represents 
the form factor which is taken to be the same as the isoscalar
nucleon electromagnetic form factor, therefore it has the following 
expression: 
\begin{eqnarray}
f(t)= F_1(t)= \frac{(4M^2_N-2.8t)}{(4M^2_N-t)(1-t/0.7)^2}.
\end{eqnarray} 

\begin{figure}[htb]
 \begin{center}
\hspace*{-10mm}  \mbox{\epsfig{file=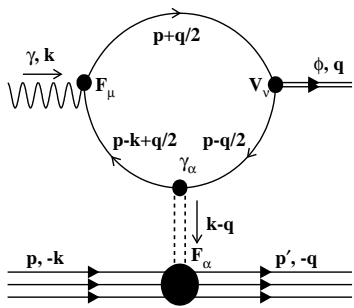,height=7.0cm}}
 \end{center}
\vspace{-3mm}
\caption{Pomeron exchange diagram in the $\phi$ meson photoproduction. 
}

\protect\label{fig:Pom}
\end{figure}

For the $\gamma\phi {\mathcal P}$ vertex, the lowest order diagram
for the quark pair creation in Ref.~\cite{lee} is used for the 
$s\overline{s}$ creation, but has been extrapolated to the 
limit of $Q^2=0$, namely, the process with real photons. 
A bare photon vertex is introduced for the quark-photon interaction, 
which has the same form as the quark-photon coupling in Eq.~(\ref{3.0}).
Here, we use the ``on-shell approximation" for the 
quark-$\phi$ vertex, i.e. $V_\nu$ in Fig.~\ref{fig:Pom}: 
\begin{eqnarray}
V_\nu(p-\frac 12 q, p+\frac 12 q)=f_\phi M_\phi\gamma_\nu \ ,
\end{eqnarray}
where $f_\phi$ represents the coupling strength and is fixed 
by the $\phi\to e^+e^-$ decay width $\Gamma_\phi\to e^+e^-$ with the 
following relation:
\begin{eqnarray}
\label{phidecay}
\Gamma_{\phi\to e^+e^-}=\frac{8\pi \alpha^2_e e^2_Q}{3}
(\frac{f^2_\phi}{M_\phi}) \ ,
\end{eqnarray}
where $e_Q=1/3$ is the charge factor of the $s$ quark 
in terms of the charge of electron.
%
%%%%%%%%%%%%%%%%%%%%%%%% FIG 1
%%%%%%%%%%%%%%%%%

Therefore, the current matrix element can be written as:
\begin{eqnarray}
\label{pomeron}
\langle p_fm_f, q\lambda_\phi|J^\mu|p_im_i \rangle
=2\beta_0 t^{\mu\alpha\nu}(k,q)\epsilon_{\phi\nu}(q){\mathcal G_P}(s,t)
\overline{u}(p_f)F_\alpha(t)u(p_i) \ ,
\end{eqnarray}
where $u(p_i)$ and $\overline{u}(p_f)$ are the initial and final 
state Dirac spinors of the protons with four-momenta $p_i$
and $p_f$, respectively.  
$k$ and $q$ are the four-momenta of the incoming photon and 
outgoing $\phi$ meson, respectively. 
$\epsilon_{\phi\nu}$ is the polarization vector of the produced 
$\phi$ meson, and the factor 2 counts the equivalent contributions
from Pomeron-quark and Pomeron-anti-quark interactions.
${\mathcal G_P}(s,t)$ is related to the Regge trajectory of the 
Pomeron 
and has the form:
\begin{eqnarray}
{\mathcal G_P}(s,t)=-i(\alpha^\prime s)^{\alpha(t)-1} \ ,
\end{eqnarray}
where $\alpha(t)=1+\epsilon+\alpha^\prime t$ is the Regge trajectory
of the Pomeron. The form factor 
$\mu^2_0/(\mu^2_0+p^2)$ is introduced for each off-shell quark line with 
four-momentum $p$. 
In Eq.~(\ref{pomeron}), $t^{\mu\alpha\nu}(k,q)$ represents 
the loop tensor and it has the following
expression for the contributing terms:
\begin{eqnarray}
\label{gaug-1}
t^{\mu\alpha\nu}(k,q)=(k+q)^\alpha g^{\mu\nu}-2k^\nu g^{\alpha\mu} \ .
\end{eqnarray}
Note that we have used the free constituent quark propagator
$S(p)=-i/(\gamma\cdot p +m_s)$ for the strange quark with mass $m_s$.
However, this latter mass term vanishes due to the odd number of 
$\gamma$ matrices in the trace for the loop.
To preserve gauge invariance, we have adopted
the transformation given in Ref.~\cite{Titov}. 
In section 4, we detail this point to investigate the effects
from the different Pomeron exchange amplitudes due to the 
various schemes for fixing the gauge.

The above discussion shows that
such a Pomeron exchange picture is consistent with the processes
described by the effective Lagrangian in Eq.~(\ref{3.0}). 
Qualitatively, 
we suppose that the effective Lagrangian governs the coupling 
of the vector meson to the constituent $s$ and $\overline{s}$
in the Pomeron exchange. Then,
the tensor part vanishes because odd number of $\gamma$-matrix 
appears in the loop integration. Only the vector part 
couples to the point-like vector meson effectively with the coupling 
constant $a$. This result reproduces the ``on-shell approximation" 
for the $s\overline{s}$-$\phi$ vertex. With this analogy, 
the parameter $a$ in the effective Lagrangian can be expressed as:
\begin{eqnarray}
a=f_\phi/M_\phi \ .
\end{eqnarray}
>From Eq.~(\ref{phidecay}), we have $f^2_\phi/M_\phi=26.6 $ MeV, then
the value of parameter $a=0.16$ is derived.
This value is comparable with what we use for the {\it s}- and 
{\it u}-channel 
quark-$\phi$-meson coupling. 
However, such an analogy does not imply a rigorous 
constraint on the parameter $a$ since generally, the tensor coupling 
will contribute especially in the large $|t|$ region
in the {\it s}- and {\it u}-channel.
But it provides us with a consistency test between the Pomeron 
exchange picture and the {\it s}- and {\it u}-channel non-diffractive 
description. As discussed in the next section, with the constraint 
from the differential cross section, $|a|=0.15$ and 
$|b^\prime |= 0.3$ 
give reasonable estimation of the contributions from the 
effective Lagrangian. 

In the case of real photons, that is $Q^2=0$, 
 the explicit expressions for the Pomeron exchange 
for the transverse and longitudinal $\phi$ production are
\begin{eqnarray}
H_{a\lambda_V}&=&32\sqrt{2}\alpha_e\pi\beta^2_0 m_\phi f_\phi 
\nonumber\\
&&\times\frac{\mu^2_0 F_1(t)}{(M^2_\phi-t)(2\mu^2_0+M^2_\phi-t)}
(\alpha^\prime s)^{\alpha(t)-1}[M_T({\mathcal P})+M_L
({\mathcal P})]_{a\lambda_V} \ ,
\end{eqnarray}
where
\begin{eqnarray}
M_T({\mathcal P})&=& -\Bigl[(\omega_\gamma+\omega_\phi)+
\frac{{\bf q}^2}{E_f+M_N}
+\frac{{\bf k}^2}{E_i+M_N} \nonumber\\
&&+(\frac{1}{E_f+M_N}+\frac{1}{E_i+M_N})
{\bf q}\cdot{\bf k}\Bigr]
{\veps \hskip -0.16 cm }_\gamma\cdot{\veps \hskip -0.16 cm }_\phi 
\nonumber\\
&&+(\frac{1}{E_f+M_N}+\frac{1}{E_i+M_N})i\vsig\cdot({\bf k}
\times{\bf q})
{\veps \hskip -0.16 cm }_\gamma\cdot{\veps \hskip -0.16 cm }_\phi \nonumber\\
&&+\frac{2}{E_f+M_N}{\veps \hskip -0.16 cm }_\phi\cdot{\bf k}
{\veps \hskip -0.16 cm }_\gamma\cdot{\bf q} \nonumber\\
&&-\frac{2}{E_f+M_N}i\vsig\cdot({\veps \hskip -0.16 cm }_\gamma\times{\bf q})
{\bf k}\cdot{\veps \hskip -0.16 cm }_\phi \nonumber\\
&&+\frac{2}{E_i+M_N}i\vsig\cdot({\veps \hskip -0.16 cm }_\gamma\times{\bf k})
{\bf k}\cdot{\veps \hskip -0.16 cm }_\phi \ ,
\end{eqnarray}
and
\begin{eqnarray}
M_\phi|{\bf q}|\times M_L({\mathcal P})&=& 
\Bigl[ -\omega_\phi(\omega_\gamma+\omega_\phi)
-\omega_\phi(\frac{{\bf q}^2}{E_f+M_N}+\frac{{\bf k}^2}{E_i+M_N})
\nonumber\\
&&-2\frac{\omega_\gamma|{\bf q}|^2}{E_f+M_N}
+\omega_\phi(\frac{1}{E_f+M_N}-\frac{1}{E_i+M_N})
{\bf q}\cdot{\bf k} \Bigr]
 {\bf q}\cdot{\veps \hskip -0.16 cm }_\gamma \nonumber\\
&& -(\omega_\gamma|{\bf q}|^2-\omega_\phi{\bf q}\cdot{\bf k})
\frac{2}{E_i+M_N}i\vsig\cdot({\veps \hskip -0.16 cm }_\gamma\times{\bf k})\nonumber\\
&& +(\omega_\gamma|{\bf q}|^2-\omega_\phi{\bf q}\cdot{\bf k})
\frac{2}{E_f+M_N}i\vsig\cdot({\veps \hskip -0.16 cm }_\gamma\times{\bf q})\nonumber\\
&& +\omega_\phi(\frac{1}{E_f+M_N}+\frac{1}{E_i+M_N})
i\vsig\cdot({\bf k}\times{\bf q}){\bf q}\cdot
{\veps \hskip -0.16 cm }_\gamma .
\end{eqnarray}
The subscripts $a$ (=1,2,3,4) and $\lambda_V$(=0,$\pm 1$) 
denote the helicity elements of the amplitude in the helicity space. 

It turns out to be impossible to extrapolate the theory of Pomeron 
exchange from high energy regions ($E_\gamma>>$ 10 GeV) down to the 
 regions 3.0$<E_\gamma<$6.7 GeV~\cite{phi1978}, 
with the same normalization factor $\beta_0$.
The cross sections will 
be over-estimated if the same normalization factor is used 
in the low energy region. 
Thus, another normalization factor must be taken for the 
quark-Pomeron 
coupling $\beta_s$ at the nucleon-Pomeron vertex.
We derive the normalization factor 
by fitting the data from Ref.~\cite{phi1978} at 
$E_\gamma$=6.45 GeV. 
It gives the {\it s}-quark-Pomeron coupling strength 
$\beta_s$=1.27 GeV$^{-1}$ (for instance, 
in Ref.\cite{lee} $\beta_s=1.5$ GeV$^{-1}$ has been adopted). 
Qualitatively, the change of the 
normalization factor can be explained by the 
non-perturbative dressing of the quark-gluon vertex required by the 
Slavnov-Taylor identity, which introduces a nontrivial flavor
dependence at the vertex~\cite{lee}. In other words, we have made the 
assumption 
that the contribution from the flavor dependence of the quark-Pomeron
vertex can be absorbed into the constant 
$\beta_s$(=$\beta_0$)~\cite{Donnachie,lee}. 
Concerning the other parameters,
we adopt the same values as used in Ref.~\cite{lee}:
$\epsilon$ = 0.08,~$\alpha^\prime$ = 0.25 GeV$^{-2}$,~$\mu_0$ = 1.2 GeV.
%
%  %%  %%  %%  %%  %%  %%  %% pi0 exchange
%
\subsection{{\it t}-channel $\pi^0$ exchange }
The Lagrangian for the
$\pi^0$ exchange has the following form:
\begin{eqnarray}\label{3}
L_{\pi NN}=-i g_{\pi NN}\overline\psi \gamma_5(\vtau\cdot\vpi)\psi
\end{eqnarray}
for the $\pi NN$ coupling vertex, and
\begin{eqnarray}\label{4}
L_{\phi \pi^0 \gamma}=e_N\frac{ g_{\phi\pi\gamma} }{M_\phi}
\epsilon_{\alpha\beta\gamma\delta}\partial^\alpha A^\beta
\partial^\gamma\phi^\delta\pi^0
\end{eqnarray}
for the $\phi\pi\gamma$ coupling vertex, where the $\phi^\delta$ 
and $\pi^0$  represent
 the $\phi$ and $\pi^0$ fields, respectively,  
 $A^\beta$ denotes the electromagnetic field,   
$\epsilon_{\alpha\beta\gamma\delta}$ is the Levi-Civita tensor, 
and $M_\phi$=1.02 GeV is the mass of the $\phi$ meson. 
The $ g_{\pi NN}$ and $ g_{\phi\pi\gamma}$ in Eqs. 
(\ref{3}) and (\ref{4}) denote the coupling constants at the two
 vertices, respectively. Therefore, 
the transition amplitudes of the {\it t}-channel $\pi^0$ 
exchange have the following expression:
\begin{eqnarray}
M_T(\pi^0)&=& \frac{e_Ng_{\pi NN} g_{\phi\pi\gamma}}{2M_\phi(t-m^2_\pi)}
\Bigl[\omega_\gamma{\veps \hskip -0.16 cm }_\gamma\cdot({\bf q}\times
{\veps \hskip -0.16 cm }_\phi)
+\omega_\phi{\bf k}\cdot({\veps \hskip -0.16 cm }_\gamma\times
{\veps \hskip -0.16 cm }_\phi)\Bigr] \nonumber \\
&&\vsig\cdot {\bf A}
e^{-\frac {({\bf q}-{\bf k})^2}{6\alpha_\pi^2}}
\label{t}
\end{eqnarray}
for the transverse transition, and
\begin{eqnarray}
M_L(\pi^0)= -\frac{e_Ng_{\pi NN} g_{\phi\pi\gamma}}{2M_\phi(t-m^2_\pi)}
\frac{ M_\phi}{|{\bf q}|}({\veps \hskip -0.16 cm }_\gamma
\times{\bf k})\cdot{\bf q} \vsig\cdot {\bf A}
e^{-\frac {({\bf q}-{\bf k})^2}{6\alpha_\pi^2}}
\label{l}
\end{eqnarray}
for the longitudinal transition, where
$\omega_\gamma$ in the transition amplitudes
denotes the energy of the photon with momentum ${\bf k}$, and the vector
${\bf A}=-\frac{{\bf q}}{E_f+M_N}+\frac{{\bf k}}{E_i+M_N}$.
The momentum transfer squared is $t=(q-k)^2=M_\phi^2-2k\cdot q$.
The factor $e^{-\frac {({\bf q}-{\bf k})^2}{6\alpha_\pi^2}}$ in Eqs. 
(\ref{t}) and (\ref{l}) plays a role like 
the form factor for both $\pi NN$ and 
$\phi \gamma \pi$ vertices. It comes out naturally
in the harmonic oscillator basis since the nucleon 
is treated as a 3-quark system which is non-point-like.
Therefore, the expansion of the internal motion gives such a 
momentum-dependent factor. 
The constant 
$\alpha_\pi$ in this form factor is treated as a parameter. 
Since $\alpha_\pi$ describes the combined form factor 
for both $\pi NN$ and 
$\phi \gamma \pi$ vertices, we do not expect it has the same value
as $\alpha=410$ MeV which only corresponds to the $\phi NN$
vertex in the {\it s}- or {\it u}-channel. 
We adopt the following values  
for the couplings:
$$ 
{g^2_{\pi NN}}/{4\pi}= 14,
~g^2_{\phi\pi\gamma}=0.143  \ ,
~\alpha_\pi = 300 \ \mbox{MeV}.
$$

It is worth noting that the inclusion of the {\it t}-channel 
$\pi^0$ exchange might result in a double-counting problem due to 
duality arguments. 
However, in the following sections, 
one can see that the $\pi^0$ exchange plays a quite negligible 
role in the $\phi$ meson photoproduction, which suggests that 
the duality hypothesis gives little constraint on this process. 
Since the duality problem is beyond a phenomenological study, 
we present results with or without the $\pi^0$ exchange to illustrate 
the effects in the following studies.

\newpage
%
%%%%%%%%%%%%%%%%%%%%%%%%%%%%%%%%%%%%% 	PHI ; OBSERVABLES AND DISCUSSION :
%
\section{Observables and discussion}
In this work, we limit the discussion to  the low energy region  
where the effects from nucleon resonances are expected to play a role in the 
reaction mechanism of the $\gamma p\to\phi p$ process.
%
%%%%%%%%%%%%%%%%%%%%%%%%%%%%%%%%%%%%% 	X-SECTION
%
\subsection{Cross-section}
As discussed in subsection 2.2, the picture
 of quark-$\phi$-meson couplings is consistent with each other  
in the Pomeron exchange and in the {\it s}- and {\it u}-channel 
mechanisms. The analogy between the above two vertices
leads to the value $a=0.16$. However,
this approach does not put any constraint on
the parameter $b^\prime$ ($\equiv b-a$), which contributes in the 
large $|t|$ region in the {\it s}- and {\it u}-channel. 

We have attempted to extract the values of 
the $a$ and $b^\prime$
parameters by fitting the differential cross section data. 
The result is : 
$$a = -0.035 \pm 0.166~;~ b^\prime = -0.338 \pm 0.075 \ . $$
First, let us mention that 
 this result should be taken with caution since we fit 
the data at forward angles whereas the influence of $a$ and $b^\prime$ 
is at large angles. However, this result shows that parameter $b^\prime$ 
is well constrained while the constraint on parameter $a$ is loose.
For $a$, rather than a precise value, the result provides a range  which 
is consistent with the value reported in
Section 2.2. In the {\it s}- and {\it u}-channel, 
the parameter $a$ reflects the vector coupling of $\phi$-$uu$ or $\phi$-$dd$ 
which should be suppressed by the OZI rule. In the above fitting, 
the small central value of $|a|=0.035$ 
shows some hints from such a suppression. 
However, recalling that one of our motivations is to investigate 
the sensitivities of the polarization observables to the small 
{\it s}- and {\it u}-channel
contributions, we have to take into account the large uncertainty 
in parameter $a$.
In other words, larger value for $a$ is to give the upper limit 
of the sensitivities of the polarization observables.
For $b^\prime$, the result favors a negative sign.
Also, with the same motivation, we will present 
in the following the results for all the phase sets for $a$ and $b^\prime$ to 
provide a complete and systematic understanding of the role 
played by the {\it s}- and {\it u}-channel contributions.

%%%%%%%%%%%%%%%%%%%%%%%% FIG 2
\begin{figure}[hb]
%% \vspace{6cm}
\begin{center}
\vspace*{-10mm}  \epsfig{file=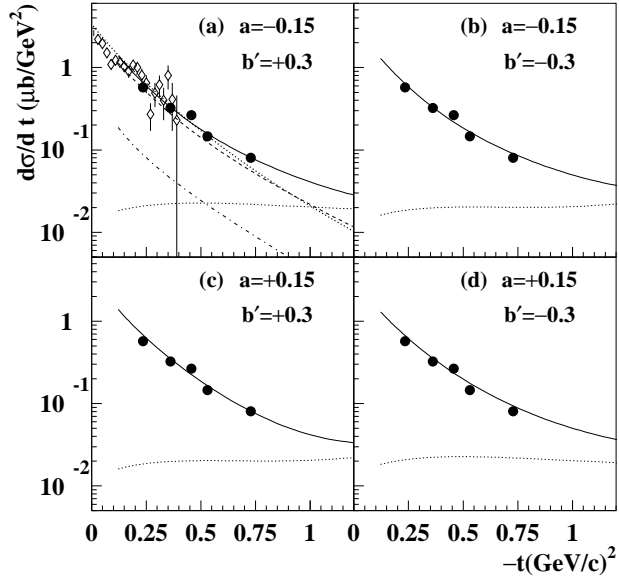,height=9.0cm,width=9.0cm}
 \end{center}
\vspace{-3mm}
\caption{
Differential cross section at $E_\gamma$ = 2.0~GeV as a function of 
momentum transfer ($-t$) for $\gamma p\to\phi p$.
Data at $E_\gamma$ = 2.0~GeV (full circle) are from Ref.~\cite{phidata},
and at $E_\gamma$ = 6.45~GeV (diamond) from Ref.~\cite{phi1978}.
The curves are: i) $\pi^0$-exchange 
(dot-dashed); ii) {\it s}- and {\it u}-channel contribution (dotted);
iii) Pomeron exchange(dashed), and iv) contributions from i) to iii) 
(full curves). 
The Pomeron exchange at $E_\gamma$ = 6.45~GeV is also depicted 
(heavy dotted curve in (a)). 
Contributions from the Pomeron and $\pi^0$ exchanges (independent
of $a$ and $b^\prime$) are only shown in (a). 
}

\protect\label{fig:ds}
\end{figure}
%%%%%%%%%%%%%%%%%

We hence fix the absolute values of the two parameters at
the following values compatible with the above ranges:
\begin{eqnarray}
\label{param}
|a|=0.15 \ ,~|b^\prime | = 0.3 \ .
\end{eqnarray}
As we will see below, these values allow to reproduce well enough the
existing data on the cross section as well as the density 
matrix elements.

Notice that: 
\begin{itemize}
{\item The extreme value $|a|=0.15$ with different signs will show
the maximum sensitivities of 
various observables to the resonance contributions. 
The OZI suppression considerations are discussed in section 3.5.}
{\item 
Since we always have the combination of the parameters $b-a$ 
in the amplitudes, we define $b^\prime\equiv b-a$, and use 
$a$ and $b^\prime$ as the two parameters for the quark-vector-meson 
coupling. }
{\item
Once the signs of the two parameters
$a$ and $b^\prime$ are determined, the phases 
between the {\it s}-, {\it u}-channel amplitudes, 
Pomeron, and $\pi^0$ exchanges can be fixed.} 
{\item 
The above values, Eq.~(\ref{param}), are smaller than those estimated 
in Ref.~\cite{prc98}. The reason is that no Pomeron 
exchange was included there, therefore, the contributions from the 
non-diffractive effective Lagrangian were over-estimated.}
\end{itemize}
In Fig.~\ref{fig:ds}, the results of our calculations 
for the differential cross section at $E_\gamma = 2.0$ GeV are
shown.
The $\pi^0$-exchange contribution is small and limited to low
momentum transfers. The {\it s}- and {\it u}- channel resonances
produce an almost flat contribution which becomes dominant roughly
above $|t| = 1.$ (GeV/c)$^2$. For the Pomeron exchange, we show results
at two energies (2.0 and 6.45 GeV). 
The Pomeron exchange is the dominant
contribution for $-t \leq 0.5$ (GeV/c)$^2$, 
and shows a soft dependence on the energy. 
As seen in Figs.~\ref{fig:ds}(b)-(d), 
the differential cross section shows basically
no sensitivity to the signs of the couplings $a$ and $b^\prime$.

%
%%%%%%%%%%%%%%%%%%%%%%%% FIG 3
\begin{figure}[htb]
%% \vspace{6cm}
\begin{center}
\hspace*{-10mm}  \mbox{\epsfig{file=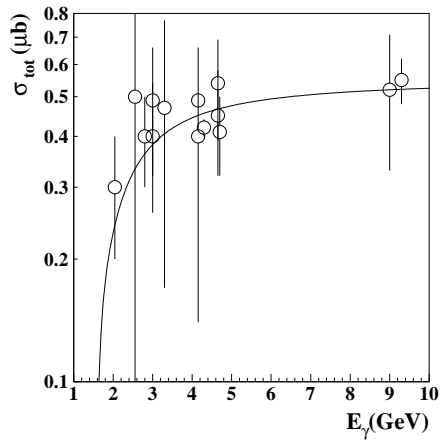,height=7.0cm}}
 \end{center}
\vspace{-3mm}
\caption{The total cross section for $\gamma p\to\phi p$.
The curve is the prediction of our full calculation and the data  
are from Ref.~\cite{phitotal}. 
}

\protect\label{fig:Tot}
\end{figure}
%%%%%%%%%%%%%%%%%

We also present predictions for the total cross section (Fig.~\ref{fig:Tot}). 
In spite of large uncertainties in the data, the theory-experiment comparison
shows that our model reproduces consistently the dominating diffractive 
 contributions in the whole phase space 
spanned by the data. Moreover, the almost energy-independent behavior
of the $\phi$ meson photoproduction process is reproduced correctly 
in the model. From these considerations, we conclude that our treatment 
of the diffractive contribution is realistic enough. As emphasized
in Introduction, based on such a reliable treatment, we can now 
investigate the role of the non-diffractive $\phi$ meson production 
mechanism in the $\gamma p \to\phi p$ process.

In the case of differential and total cross sections, the non-diffractive
effects turn out to be small.
However, as we will see below, such small effects can be amplified 
in polarization observables. 
The rest of this section is devoted to the polarization observable
asymmetries where the differential cross section enters in the denominator.
%
%%%%%%%%%%%%%%%%%%%%%%%%%%%%%%%%%%%%% 	SINGLE-POL
%
\subsection{Single polarization asymmetries}
\begin{figure}[htb]
\begin{center}
\hspace*{-10mm}  \mbox{\epsfig{file=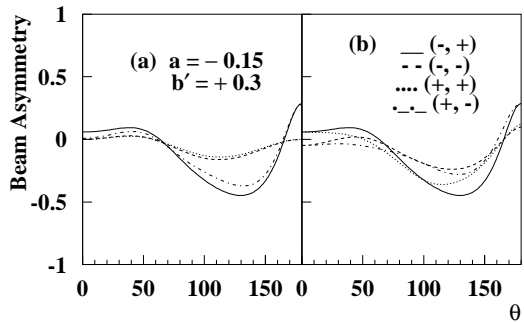,height=5.0cm,width=10.0cm}}
\end{center}
\vspace{-3mm}
\caption{The polarized beam asymmetry at $E_\gamma$ =2.0~GeV
with different phase signs. The curves in (a) stand for:
Pomeron exchange (dashed),
Pomeron and $\pi^0$ exchanges (dotted),
Pomeron exchange and resonance contributions (dot-dashed),
and the full calculation including all three components
with $a=-0.15$, $b^\prime=0.3$ (full).
In (b), the results of our full calculations for the four
($a$, $b^\prime$) sets are depicted.
}

\protect\label{fig:Beam}
\end{figure}

The beam polarization asymmetry  $ \check{\Sigma}$ at 
$E_\gamma$ = 2.0~GeV is shown in Fig.~\ref{fig:Beam}. 
Comparing the Pomeron exchange (dashed curve in Fig.~\ref{fig:Beam}~(a))
with the Pomeron plus $\pi^0$ exchange 
(dotted curve in Fig.~\ref{fig:Beam}~(a)), 
we find that the contribution from $\pi^0$ exchange is negligible.  
The {\it s}- and {\it u}-channel contributions 
amplified by the Pomeron exchange,
due to the interference terms, increase the magnitude
of the observable by about a factor of 3 around 110$^\circ$
and produces a sign change 
above 150$^\circ$ (dot-dashed curve in Fig.~\ref{fig:Beam}~(a)).
The three mechanisms together produce the full curves 
in Fig.~\ref{fig:Beam}~(a) and (b). 
We present the results for the four phase sets 
in Fig.~\ref{fig:Beam}(b) for comparison.
%%%%%%%%%%%%%%%%%%%%%%%% FIG 4
%%%%%%%%%%%%%%%%%

\begin{figure}[htb]
\begin{center}
\hspace*{-10mm}  \mbox{\epsfig{file=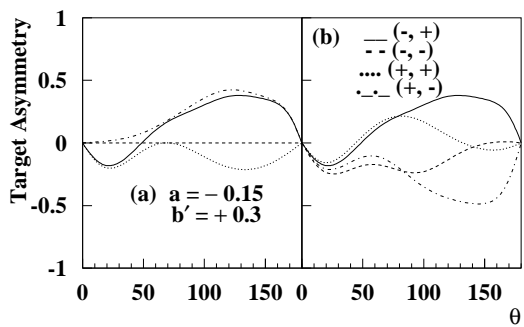,height=5.0cm,width=10.0cm}}
\end{center}
\vspace{-3mm}
\caption{Same as Fig.~\ref{fig:Beam}, 
but for the polarized target asymmetry. 
}

\protect\label{fig:Target}
\end{figure}

In Fig.~\ref{fig:Target} predictions for the target polarization asymmetry 
$\check{T}\equiv {\bf P}_N\cdot\hat{y}{\mathcal T}$, due to the same mechanisms
discussed above in the case of the $ \check{\Sigma}$ observable are
reported. 

%
%%%%%%%%%%%%%%%%%%%%%%%% FIG 5
%%%%%%%%%%%%%%%%%
%  
Here, we wish to emphasize that the helicity amplitude structure
of the $ \check{\Sigma}$ observable differs drastically from those
of the other single polarization observables. As summarized in
Appendix, the $ \check{\Sigma}$ observable is a bilinear combination
of real-real or imaginary-imaginary parts, while the other three
single polarization observables depend on real-imaginary couples.
Moreover, the Pomeron exchange amplitude is treated purely imaginary 
in this model 
while that of the $\pi^0$ exchange is purely real.
Therefore, the pure Pomeron exchange term, leads to a zero
target asymmetry (dashed curve in Fig.~\ref{fig:Target}(a)),  
while adding the $\pi^0$ exchange produces non-zero effects, 
especially at large angles (dotted curve in Fig.~\ref{fig:Target}(a)). 
We find that a large cancelation arises 
between the longitudinal and transverse parts of the asymmetry, 
which produces 
a nearly zero asymmetry at $\sim 65^\circ$. This structure 
is independent on the relative phase between the Pomeron exchange 
and the $\pi^0$ exchange amplitudes, since the Pomeron exchange 
amplitude is purely imaginary and the $\pi^0$ exchange is purely 
real, therefore, the phase change will only give an overall sign to the 
dotted curve in Fig.~\ref{fig:Target}(a). 

The Pomeron plus resonances contributions 
(dot-dashed curve in Fig.~\ref{fig:Target}(a)) gives even a larger
asymmetry in magnitude, with opposite sign for backward angles, than 
the Pomeron plus $\pi^0$ exchange does. 
The full calculation (full curves in Fig.~\ref{fig:Target}(a) and (b))
shows a minimum around 20$^\circ$ due to $\pi^0$ exchange and
a maximum around 130$^\circ$ generated by the resonance terms.
In both cases the Pomeron exchange plays an amplifying role in
the predicted asymmetries. 
It shows that the target 
polarization asymmetry is governed mainly by the resonance contributions 
at large angles. For comparison, we also present the results 
with phase changes in Fig.~\ref{fig:Target}(b). 
%%%%%%%%%%%%%%%%%%%%%%%% FIG 6
\begin{figure}[htb]
\begin{center}
\hspace*{-10mm}  \mbox{\epsfig{file=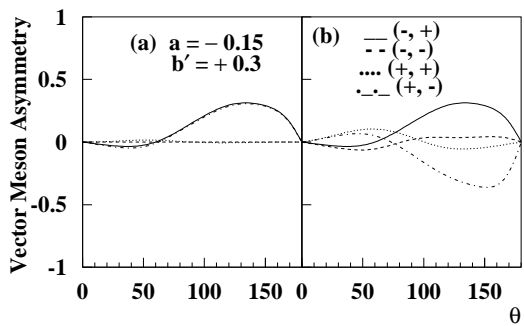,height=5.0cm,width=10.0cm}}
\end{center}
\vspace{-3mm}
\caption{Same as Fig.~\ref{fig:Beam}, 
but for the polarized vector meson asymmetry. 
}

\protect\label{fig:Pol-phi}
\end{figure}
%%%%%%%%%%%%%%%%% 
%
%%%%%%%%%%%%%%%%%%%%%%%% FIG 7
\begin{figure}[htb]
\begin{center}
\hspace*{-10mm}  \mbox{\epsfig{file=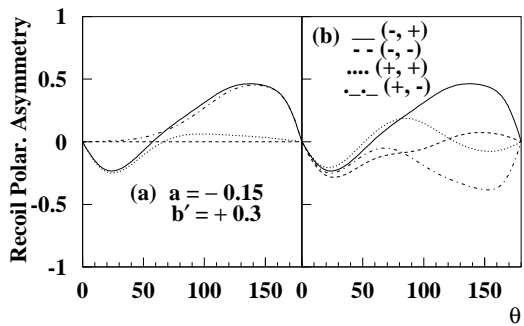,height=5.0cm,width=10.0cm}}
\end{center}
\vspace{-3mm}
\caption{Same as Fig.~\ref{fig:Beam}, 
but for the recoil polarization asymmetry. 
}

\protect\label{fig:Pol-R}
\end{figure}
%%%%%%%%%%%%%%%%%

In Fig.~\ref{fig:Pol-phi}, the vector polarization 
observable $\check{P}_V$
for the vector meson polarization
 is depicted. Here, the combined Pomeron and 
 $\pi^0$ exchanges produce a very small positive asymmetry. 
The dominant effect 
 is therefore due to the 
contributions  from the {\it s}- and {\it u}-channel resonances once 
again amplified by the Pomeron exchange 
(dot-dashed curve in Fig.~\ref{fig:Pol-phi}(a)).

Finally, we show in 
Fig.~\ref{fig:Pol-R} the recoil polarization asymmetry 
$\check{P}_{N^\prime}$. 

To summarize the main features revealed by the
single polarization observables:
\begin {itemize}
\item {The Pomeron exchange mechanism turns out to be an efficient
amplifier for the mechanisms suppressed in the cross sections.}
\item {
Although the influence of the  $\pi^0$ exchange 
can be amplified in some polarization observables, 
it plays in general a rather minor role. 
Therefore, this might imply that the double-counting from duality
(if it exits) is negligible.}
\item {The nodal structure of the observables depends (in some cases heavily)
on the signs of the two couplings $a$ and $b^\prime$.}
\item {The {\it s}- and {\it u}-channel resonances produce significant
effects. The most favorable phase space region depends on the signs
of the couplings $a$ and $b^\prime$ (Figs.~\ref{fig:Beam}(b),
 \ref{fig:Target}(b),
\ref{fig:Pol-phi}(b), and \ref{fig:Pol-R}(b)).}
\end {itemize}
%
%
%%%%%%%%%%%%%%%%%%%%%%%%%%%%%%%%%%%%% 	DOUBLE-POL
%
\subsection{Double polarization asymmetries}
Given the availability of polarized beam and polarized target,
we now concentrate on the beam-target (BT) double polarization
asymmetry. Another motivation in investigating this observable
is that a recently developed strangeness knock-out model~\cite{Titov} 
suggests that a small $s\overline{s}$ component ($\sim 5\%$) in the
proton might result in large asymmetries ($\sim 25$-$45\%$) 
in the BT observable at small angles. 
However, since the 
resonance contributions have not been taken into account there,
an interesting question is: if contributions from the {\it s}- 
and {\it u}-channel can produce a significant double polarization
asymmetry without introducing 
strangeness component or not.
%
%%%%%%%%%%%%%%%%%%%%%%%% FIG 8
\begin{figure}[htb]
\begin{center}
\hspace*{-10mm}  \mbox{\epsfig{file=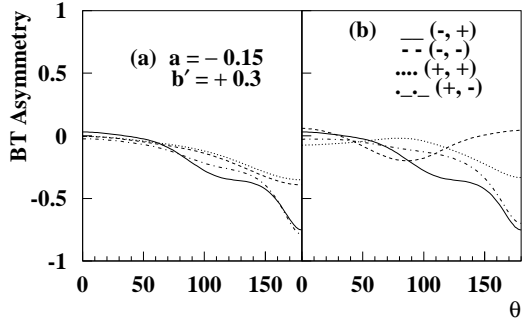,height=5.0cm,width=10.0cm}}
\end{center}
\vspace{-3mm}
\caption{Same as Fig.~\ref{fig:Beam}, 
but for the polarized beam-target asymmetry. 
The heavy dotted curve in (a) is given by the full calculation 
with $a$=-0.015, $b^\prime=+0.3$. 
}

\protect\label{fig:Pol-BT}
\end{figure}
%%%%%%%%%%

Our predictions are shown in Fig.~\ref{fig:Pol-BT}.
The Pomeron exchange alone (dashed curve in (a)), gives
a negative asymmetry which increases in magnitude from forward
to backward angles where the largest asymmetry is about 40\%.
The $\pi^0$ exchange (dotted curve in (a)) diminishes slightly
the asymmetry, while the resonances contributions (dot-dashed curve
in (a)) enhances it. The full calculation leads finally to a 
decreasing behavior, going from almost zero at 
forward angles to $ \sim -0.7$ at 180$^\circ$. This result 
(full curve in Fig.~\ref{fig:Pol-BT}(a) and (b)) is obtained
with $a=-0.15$ and $b^\prime=+0.3$. The backward angle effects are
also large in the case of $a=+0.15$ and 
$b^\prime=-0.3$ (dot-dashed curve in Fig.~\ref{fig:Pol-BT}(b)).
The situation becomes very different for the  
couplings sets with the same signs: 
the effect is suppressed for $a=+0.15$ and $b^\prime=+0.3$ 
(dotted curve in Fig.~\ref{fig:Pol-BT}(b)),
and the shape changes drastically for $a=-0.15$ and $b^\prime=-0.3$
(dashed curve in Fig.~\ref{fig:Pol-BT}(b)).
The latter set produces (almost) vanishing values at extreme angles.
The common feature to all four sets is that the beam-target
asymmetry is small at forward angles. 

%%%%%%%
%
%%%%%%%%%%%%%%%%%%%%%%%%%%%%%%%%%%%%% 	DENSITY-MATRIX
%
\subsection{Density matrix elements}
In this subsection, the density matrix elements are investigated 
in the helicity system.
Data for the density matrix elements at low energies~\cite{phi1978} 
are still very sparse:
measurements have been carried out only in the small $|t|$ region 
($\theta_{cm}< 20^\circ$) at 
$E_\gamma=5.165$, and 6.195 GeV,
where the Pomeron exchange dominates over other non-diffractive 
processes. 
%
%%%%%%%%%%%%%%%%%%%%%%%% FIG 9
\begin{figure}[htb]
\begin{center}
\hspace*{-10mm}  \mbox{\epsfig{file=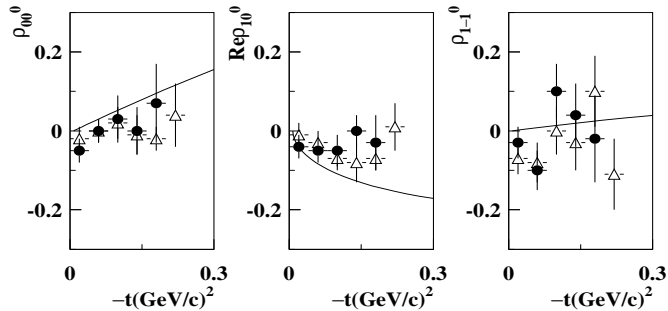,height=9.0cm,width=10.0cm}}
\end{center}
\vspace{-40mm}
\caption{Theoretical predictions for the density matrix elements at 
$E_\gamma=5.165$ GeV. Data at $E_\gamma=5.165$ GeV (full circle) 
and at $E_\gamma=6.195$ GeV (triangle) are from Ref.~\cite{phi1978}.
$-t=0.3$ (GeV/c)$^2$ corresponds to $\theta_{cm}\sim 22^\circ$. }

\protect\label{fig:Densi-1}
\end{figure}
%%%%%%%%%%%%%%%%%
In Fig.~\ref{fig:Densi-1} the solid curve is the result of our 
calculation at $E_\gamma=5.165$ GeV. 
At $E_\gamma=6.195$ GeV, the results are not significantly different
within this momentum transfer region,
so, we do not show them here. 
Theory-data comparison shows again the character of 
diffractive dominance in the small $|t|$ region, and the Pomeron exchange 
(other amplitudes are included but negligible at this energy region) 
reproduces well enough the data for the three 
density matrix elements $\rho^0_{00}$, $Re\rho^0_{10}$ and 
$\rho^0_{1-1}$.

%
%%%%%%%%%%%%%%%%%%%%%%%% FIG 10
\begin{figure}[htb]
\begin{center}
\hspace*{-10mm}  \mbox{\epsfig{file=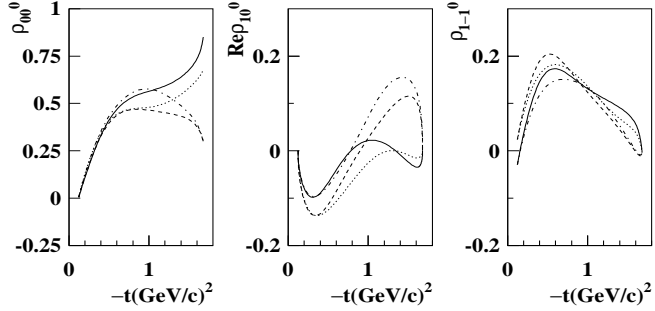,height=9.0cm,width=10.0cm}}
\end{center}
\vspace{-40mm}
\caption{ Density matrix elements predicted at $E_\gamma=2.0$ GeV 
with phase changes for $(a, \ b^\prime)$:
solid $(-, \ +)$, dashed $(-, \ -)$, dotted $(+, \ -)$, and 
dot-dashed $(+, \ +)$ curves. 
}

\protect\label{fig:Densi-2}
\end{figure}
%%%%%%%%%%%%%%%%%
%
%%%%%%%%%%%%%%%%%%%%%%%% FIG 11
\begin{figure}[htb]
\begin{center}
\hspace*{-10mm}  \mbox{\epsfig{file=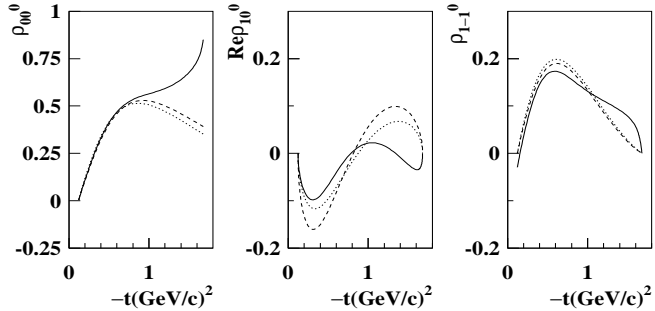,height=9.0cm,width=10.0cm}}
\end{center}
\vspace{-40mm}
\caption{ Density matrix elements predicted  
at $E_\gamma=2.0$ GeV with $ a=-0.15$, $b^\prime=0.3$. 
The curves stand for: pure Pomeron exchange (dashed),
Pomeron plus $\pi^0$ exchanges (dotted), and full calculation (full).
}

\protect\label{fig:Densi-3}
\end{figure}
%%%%%%%%%%%%%%%%%

In Fig.~\ref{fig:Densi-2}, our predictions for the three density matrix 
elements at $E_\gamma=2.0$ GeV are presented for different 
$a$ and $b^\prime$ phase sets.
In Fig.~\ref{fig:Densi-3}, 
contributions from only the Pomeron exchange (dashed curve), 
Pomeron plus $\pi^0$ exchanges (dotted curve),
and both exchanges plus the resonance contributions (full curve)
are shown. 
The density matrix elements appear to be quite sensitive 
to the non-diffractive {\it s}- and {\it u}-channel contributions. 
Also, the phase changes
produce significant effects in the large $|t|$ region. 
However, in the forward direction, the matrix elements
are not sensitive neither to the non-diffractive resonance effects nor
to the phase changes. 
\subsection{OZI suppression effects}
As mentioned in subsection 3.1, the numerical results discussed there
are obtained for $|a|= 0.15$, which corresponds to no OZI suppression.
Here, we report on the manifestations of such a suppression
in the absence of OZI evading mechanisms. The OZI rule would imply a
parameter value smaller by roughly one order of magnitude. To single
out possible manifestations of the OZI suppression, we compare 
the numerical results for different observables for $|a|= 0.15$ and
$a= 0.015$. 

As already discussed, from the Pomeron-photon
analogy picture, no constraint can be imposed
on the other free parameter, $b^\prime$. However, the value
$|b^\prime|$=0.3 
could indicate an OZI suppression: the value extracted here is about one 
order of magnitude smaller than  $b^\prime= 2.5$ used in 
the $\omega$ and $\rho$ mesons photoproduction~\cite{prc98}.

\begin{figure}[htb]
\begin{center}
\hspace*{-10mm}  \mbox{\epsfig{file=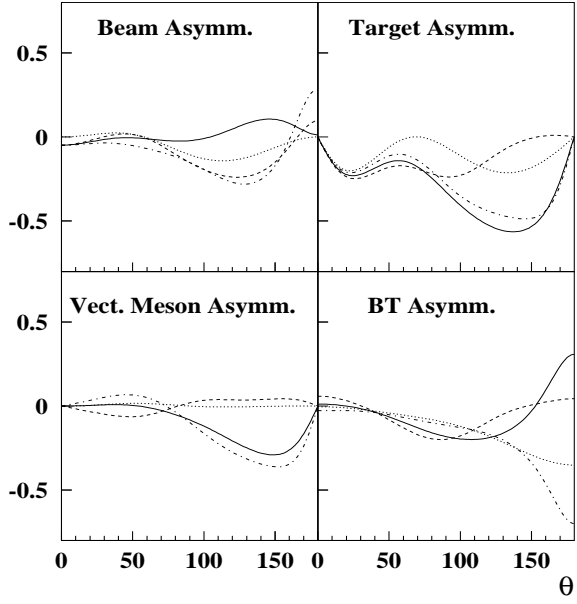,height=9.0cm,width=10.0cm}}
\end{center}
\vspace{-3mm}
\caption{ Single and double polarization observables predicted  
at $E_\gamma=2.0$ GeV with $b^\prime=-0.3$. 
The curves stand for:  
Pomeron plus $\pi^0$ exchanges (dotted), and full calculation for
$a=0.15$ (dot-dashed), $a= -0.15$ (dashed), and $a = 0.015$ (solid).}
\protect\label{fig:OZI-1}
\end{figure}

In Fig.~\ref{fig:OZI-1}, our results for three single polarization
(beam, target, and vector meson), as well as the
beam-target double polarization asymmetries are depicted. 
The polarized beam asymmetry shows little sensitivity to the phase of
$a$, but depends significantly on the absolute value of this parameter. This
observable is hence very appealing to study the OZI suppression 
effects and/or the related evading mechanisms. 
In the target and vector meson polarization
asymmetries, the $a$-dependent terms come in basically in the 
interference terms and are mostly sensitive to the phase of this
parameter. 
The beam-target double polarization asymmetry, 
turns out to result from cancellations among the helicity amplitudes,
except at extreme backward angles where strong dependences on both the
phase and magnitude of $a$ show up.
This explains the 
large asymmetries (solid curve in Fig.~\ref{fig:OZI-1})
found for a small value of the coupling $a$. 
The curves without {\it s}- and {\it u}-channel contributions 
(i.e. only Pomeron plus $\pi^0$ exchange) are also shown for comparison. 

\begin{figure}[htb]
\begin{center}
\hspace*{-10mm}  \mbox{\epsfig{file=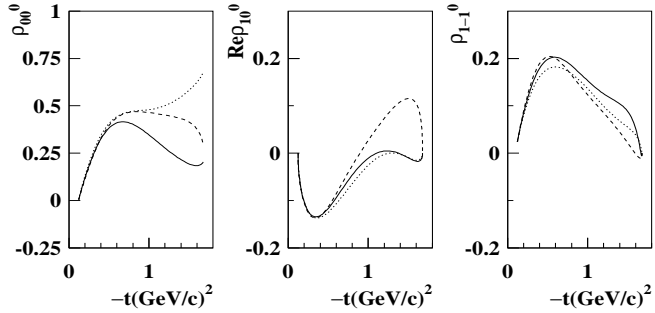,height=9.0cm,width=10.0cm}}
\end{center}
\vspace{-40mm}
\caption{Density matrix elements predicted  
at $E_\gamma=2.0$ GeV with $b^\prime=-0.3$. 
The curves stand for: full calculation for
$a=0.15$ (dotted), $a= -0.15$ (dashed), and $a= 0.015$ (solid).}
\protect\label{fig:OZI-2}
\end{figure}

Density matrix elements are shown in Fig.~\ref{fig:OZI-2}. 
The $\rho^0_{00}$ above $|t| \approx$ 1. (GeV/c)$^2$ shows a
possibility to determine both the phase and the size of
the $a$ parameter, while the $Re\rho^0_{10}$ depends strongly
on the phase. The $\rho^0_{1-1}$ turns out to be dominated by the 
non-resonant terms (see Fig.~\ref{fig:Densi-3}).

The results presented here show clear sensitivity of some of the
observables to the nucleon resonance contributions even in the
presence of the OZI suppression mechanism.

%%%%%%%%%%%%%%%%%%%%%%%% FIG 12
%%%%%%%%%%%%%%%%%
\vspace{-20mm}
%%%%%%%%%%%%%%%%%%%%%%%% FIG 13
%%%%%%%%%%%%%%%%%%
\clearpage
\section{Pomeron gauge invariance effects}
This section is devoted to investigate the sensitivity of the Pomeron 
exchange model to the various gauge-fixing schemes. This question arises from 
the fact that the Pomeron exchange is introduced phenomenologically 
to account for the diffractive behavior at small angles. 
However, at large angles, the description of the Pomeron structure  
is still an unsolved question.
Here, we limit our discussion to the effects due to 
the different gauge-fixing 
schemes and the non-gauge-invariant one.

In Fig.\ref{fig:Pom}, the loop tensor is derived by supposing that
the constituent quarks carry half of the momentum of the vector meson: 
\begin{eqnarray}
t^{\mu\alpha\nu}(k,q)&=&2k^\alpha g^{\mu\nu} -\frac{2}{q^2}k^\alpha
q^\mu q^\nu -2g^{\alpha\mu}(k^\nu-q^\nu\frac{k\cdot q}{q^2})\nonumber\\
&& +2(k^\mu-q^\mu)(g^{\nu\alpha}-\frac{q^\alpha q^\nu}{q^2}) \ ,
\end{eqnarray}
where the second line violates gauge invariance. Taking into 
account that at the photon and vector meson vertices,
$k\cdot \epsilon_\gamma=0$ and $q\cdot \epsilon_\phi=0$,
one finds that the contributing terms are: 
\begin{equation}
t^{\mu\alpha\nu}(k,q)=2k^\alpha g^{\mu\nu}-2g^{\alpha\mu} k^\nu 
-2q^\mu g^{\nu\alpha} \ .
\end{equation}
We note the above non-gauge-invariant Pomeron amplitude 
as {\bf t0}. In Ref.\cite{Titov}, two schemes for restoring 
gauge invariance are employed. 
We refer reader to Ref.\cite{Titov} for detailed discussion of 
how to derive the 
gauge-fixing terms, and skip to the final forms of the 
loop tensor. 

In this work 
we have adopted the gauge-fixing scheme  
which gives the loop tensor as the following: 
\begin{eqnarray}
t^{\mu\alpha\nu}(k,q)&=&(k+q)^\alpha g^{\mu\nu}
 -2k^\nu g^{\alpha\mu}\nonumber\\
&& +2\Bigl[ k^\mu g^{\alpha\nu}+\frac{q^\nu}{q^2}(k\cdot q g^{\alpha\mu}
-k^\alpha q^\mu -q^\alpha k^\mu) \nonumber\\
&& -\frac{k^2 q^\mu}{q^2 k\cdot q}(q^2 g^{\alpha\nu} -q^\alpha q^\nu)\Bigr]
 +(k-q)^\alpha g^{\mu\nu} \ .
\end{eqnarray}
The contributing terms can be easily derived 
and are given by Eq.~(\ref{gaug-1}). Here we name this gauge-fixing
scheme as {\bf t1}.

Another gauge fixing scheme gives the contributing terms as: 
\begin{equation}
t^{\mu\alpha\nu}(k,q)=(k+q)^\alpha g^{\mu\nu}-k^\nu g^{\alpha\mu}
-q^\mu g^{\alpha\nu} \ ,
\end{equation}
where those terms which do not contribute are omitted for brevity.
We note this gauge-fixing scheme as {\bf t2}.

Comparing these three kinds of Pomeron exchanges: 
{\bf t0}, non-gauge-invariant; {\bf t1} and {\bf t2}, gauge-invariant,
one finds that these different Pomeron exchanges will produce different 
behaviors. However, one can identify the two terms which are different 
between the three Pomeron amplitudes: 
$q\cdot\epsilon_\gamma \gamma\cdot\epsilon_\phi$
and $q\cdot \gamma \epsilon_\gamma\cdot\epsilon_\phi$, and finds 
that they generally play a role as a higher order contribution
at forward angles.
For $q\cdot\epsilon_\gamma \gamma\cdot\epsilon_\phi$ 
we write explicitly the non-relativistic
expansion as follows:
\begin{eqnarray}
M_T & = & -{\bf q}\cdot {\veps \hskip -0.16 cm }_\gamma 
\frac{{\bf k}\cdot{\veps \hskip -0.16cm}_\phi}{E_i+M_N}\nonumber\\
&&+{\bf q}\cdot {\veps \hskip -0.16 cm }_\gamma
\left[ \frac{i\vsig\cdot({\veps \hskip -0.16cm}_\phi\times{\bf q})}{E_f+M_N}
-\frac{i\vsig\cdot({\veps \hskip -0.16cm}_\phi\times{\bf k})}{E_i+M_N}
\right] \ ,
\end{eqnarray}
and
\begin{eqnarray}
\label{gauge-term1}
M_\phi |{\bf q}|\times M_L & = & -{\bf q}\cdot {\veps \hskip -0.16 cm }_\gamma
\left( |{\bf q}|^2 +\omega_\phi (\frac{{\bf q}^2}{E_f+M_N}
+\frac{{\bf q}\cdot{\bf k}}{E_i+M_N})\right) \nonumber\\
&&- \omega_\phi {\bf q}\cdot {\veps \hskip -0.16 cm }_\gamma
\frac{i\vsig\cdot({\bf q}\times{\bf k}) }{E_i+M_N} \ ,
\end{eqnarray}
where $M_T$ and $M_L$ represent the transverse and longitudinal 
transition amplitude, respectively. 
Also, this term is the one which makes the difference between 
 the two gauge-fixing 
schemes, {\bf t1} and {\bf t2}. 
 However, at forward angles, this term becomes small and vanishes 
since ${\bf q}$ has almost the same direction as ${\bf k}$ 
in the forward scattering, therefore,
the product ${\bf q}\cdot {\veps \hskip -0.16 cm }_\gamma$ becomes 
very small.  Qualitatively, we can see that the product of 
${\bf q}\cdot {\veps \hskip -0.16 cm }_\gamma$ gives an overall
suppression of this term, which guarantees the consistent behavior
of the Pomeron exchanges at small angles with different gauge-fixing 
schemes.

For the second term, $q\cdot \gamma \epsilon_\gamma\cdot\epsilon_\phi$, 
which is also essential to restore gauge invariance, 
we find that there is also an overall suppression 
of the longitudinal amplitude from 
${\bf q}\cdot {\veps \hskip -0.16 cm }_\gamma$ in the forward direction:
\begin{eqnarray}
\label{gauge-term2}
M_\phi |{\bf q}|\times M_L & = & -{\bf q}\cdot {\veps \hskip -0.16 cm }_\gamma
\left( \omega^2_\phi +\omega_\phi (\frac{{\bf q}^2}{E_f+M_N}
+\frac{{\bf q}\cdot{\bf k}}{E_i+M_N})\right) \nonumber\\
&&- \omega_\phi {\bf q}\cdot {\veps \hskip -0.16 cm }_\gamma
\frac{i\vsig\cdot({\bf q}\times{\bf k}) }{E_i+M_N} \ .
\end{eqnarray}
Note that in Eqs.~(\ref{gauge-term1}) and (\ref{gauge-term2}), 
except for the first term in the first line, 
the other terms are identical. 
Therefore, the substitution of Eq.~(\ref{gauge-term1})
and (\ref{gauge-term2}) into scheme {\bf t2} in
the longitudinal amplitude for the two terms gives, 
$M_L=-M_\phi {\bf q}\cdot {\veps \hskip -0.16 cm }_\gamma /|{\bf q}| $, 
which is obviously suppressed in the forward direction.

%%%%%%%%%%%%%%%%%%%%%%%% FIG 14
\begin{figure}[htb]
%\vspace{50mm}
\begin{center}
\hspace*{-10mm}  \mbox{\epsfig{file=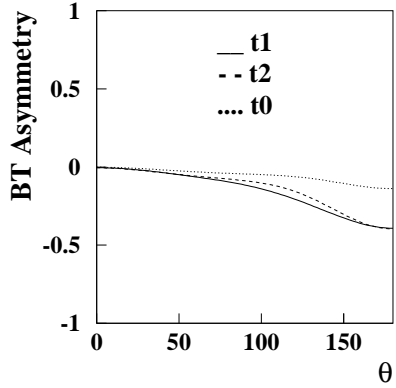,height=7.0cm,width=7.0cm}}
\end{center}
\vspace{-3mm}
\caption{ 
Beam-target asymmetry for different Pomeron exchange amplitudes. 
The non-diffractive 
contributions have been switched off.}

\protect\label{fig:gaug_BT1}
\end{figure}
%
%%%%%%%%%%%%%%%%%%%%%%%%%%%%
 
For the transverse amplitude of the
$q\cdot \gamma \epsilon_\gamma\cdot\epsilon_\phi$ term, 
the explicit expression  
is:
\begin{eqnarray}
M_T&=& -(\omega_\phi +\frac{{\bf q}^2}{E_f+M_N}
+\frac{{\bf q}\cdot{\bf k}}{E_i+M_N})
{\veps \hskip -0.16 cm }_\gamma\cdot{\veps \hskip -0.16 cm }_\phi
\nonumber\\
&&-\frac{i\vsig\cdot({\bf q}\times{\bf k})}{E_i+M_N} 
{\veps \hskip -0.16 cm }_\gamma\cdot{\veps \hskip -0.16 cm }_\phi \ ,
\end{eqnarray} 
where the second term is suppressed by $({\bf q}\times {\bf k})$ 
in the forward direction. However, for the first term in the above 
equation, the only suppression comes from the relatively smaller 
momentum of the massive $\phi$ meson. 

%%%%%%%%%%%%%%%%%%%%%%%% FIG 15
\begin{figure}[htb]
%\vspace{50mm}
\begin{center}
\hspace*{-10mm}  \mbox{\epsfig{file=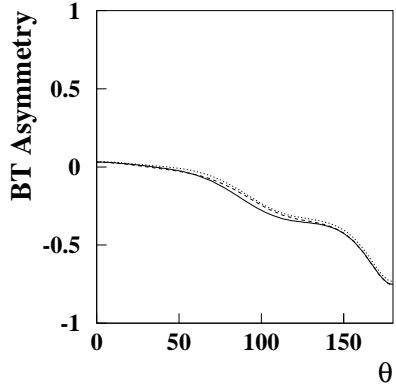,height=7.0cm,width=7.0cm}}
\end{center}
\vspace{-3mm}
\caption{ Beam-target asymmetries with all the contributions 
taken into account. The Pomeron exchange is given by {\bf t1} (full), 
{\bf t2} (dashed), and {\bf t0} (dotted). $a=-0.15$, $b^\prime=0.3$ 
have been adopted for the {\it s}- and {\it u}-channel contributions.
}

\protect\label{fig:gaug_BT2}
\end{figure}
%
%%%%%%%%%%%%%%%%%%%%%%%%%%%%

%
%%%%%%%%%%%%%%%%%%%%%%%% FIG 16
\begin{figure}[htb]
%\vspace{50mm}
\begin{center}
\hspace*{-10mm}  \mbox{\epsfig{file=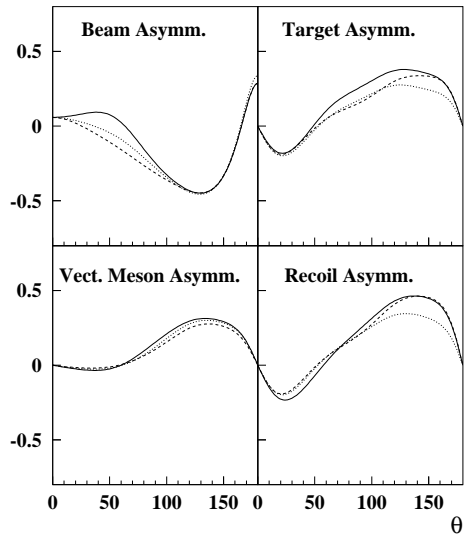,height=8.0cm,width=8.0cm}}
\end{center}
\vspace{-3mm}
\caption{ 
Single asymmetries with all the contributions 
taken into account. The curves have the same meanings as 
in Fig.~\ref{fig:gaug_BT2}. }

\protect\label{fig:gaug_single}
\end{figure}
%
%%%%%%%%%%%%%%%%%%%%%%%%%%%%

In Fig.\ref{fig:gaug_BT1}, the beam-target asymmetry for different 
Pomeron exchanges (with only the different Pomeron contribution)
 are presented. It shows that they consistently 
converge to small asymmetry at forward angles while at large angles
the asymmetry is quite sensitive to the particular Pomeron structure. The solid 
curve is given by the {\bf t1} gauge-fixing scheme 
and the dotted curve by {\bf t2}. The non-gauge-invariant Pomeron {\bf t0}
(dashed curve) 
has a similar behavior as {\bf t1}. Therefore, a question arising 
from this result is that ``if the non-diffractive contributions 
are taken into account, the polarization asymmetries are 
sensitive to the Pomeron structure or not?" 
Since the Pomeron structure 
at large angles is unknown, a gauge-dependent 
Pomeron interference at large angles might make the asymmetry 
predictions misleading. To investigate this aspect, we calculate 
the polarization asymmetries with different Pomeron exchanges
given by {\bf t0}, {\bf t1} and {\bf t2}, and the results are presented 
in Fig.~\ref{fig:gaug_BT2} and \ref{fig:gaug_single}.  
Comparing the results given by the different Pomeron exchanges, 
we find that they have very similar behaviors in the polarization 
asymmetry observables which shows that the asymmetries are not 
sensitive to the Pomeron structures when the non-diffractive 
contributions are taken into account. 
In other words, although 
different gauge-fixing schemes are introduced for the Pomeron 
exchange terms, they produce a quite gauge-independent interference 
in the polarization asymmetries, which suggests that the asymmetries can 
be considered mainly determined by the non-diffractive 
contributions. Since the asymmetries predicted at large 
angles have little dependence on the Pomeron exchange models, 
the experimental observation can provide a direct test of the model for the
{\it s}- and {\it u}-channel reactions. 

In summary, although the Pomeron exchange can have different structures 
due to the different gauge-fixing schemes, it does not influence  
significantly the polarization observables at large angles. 
Therefore, the main feature 
of an asymmetry can be regarded as depending more on the non-diffractive 
contributions rather than on the particular Pomeron structure.

\clearpage
%
%  %%  %%  %%  %%  %%  %%  %%  CONCLUSION
%
\section{ Conclusions }
In this paper we have investigated the reaction 
$\gamma p \to \phi p$ in the region close to threshold
 ($E^{thres}_{\gamma} \sim 1.57$ GeV). The known dominant diffractive
process is taken into account {\it via} the {\it t}-channel Pomeron
exchange. 
The main novelty of our work is the study of the role played
by the {\it s}- and {\it u}-channel intermediate nucleonic 
resonances. This is done using a quark-model based effective Lagrangian
approach. 
Our formalism includes also the small non-diffractive
contribution from the {\it t}-channel $\pi^0$ exchange. 

In the first step, the nucleonic resonance effects are investigated 
in the $SU(6)\otimes O(3)$ symmetry limit. In the polarization 
observables, the small nucleonic resonance contributions produce 
significant asymmetries at large angles, however, the asymmetries 
are quite small at forward angles. 

Based on the different gauge-fixing schemes for the Pomeron exchange 
terms, three kinds of Pomeron amplitudes are investigated. It shows that 
these amplitudes converge to a similar behavior at small angles, but 
have quite different behaviors at large angles when only the Pomeron 
exchange contributes. 
However, with the non-diffractive {\it s}- and {\it u}-channel
contribution taken into account, we find that the asymmetries at large 
angles are governed by the non-diffractive contributions. Conclusively, 
the asymmetries are not sensitive to the 
different Pomeron exchange amplitudes due to gauge invariance. 
This gives us the confidence that at large angles 
the ambiguity arising from the Pomeron structure can be neglected 
in this model.

The quark model presented in this work is based on the
$SU(6)\otimes O(3)$ symmetry. However, the breaking of this
symmetry will
lead to the configuration mixings among states with
the same quantum numbers. This mechanism allows those
resonances excluded by virtue of the $SU(6)\otimes O(3)$ symmetry
to contribute in this reaction and suppresses
the role played by other members of the degenerate states.
The present data on the $\phi$ meson photoproduction are too scarce
to allow a close study of the possible deviations 
from the exact $SU(6)\otimes O(3)$ symmetry. 
As a rough estimation of such an effect,
we studied the role of individual resonances
 by including or eliminating their contributions in the polarization
observables. 
We find that most of the single and double polarization
observables investigated here show significant effects only
at large angles.

In summary, the forward angle polarization asymmetries are 
almost insensitive to the {\it s}- and {\it u}-channel nucleonic 
resonance contributions
(as well as to the {\it t}-channel $\pi^0$ exchange). 
This is an interesting finding, since data at small angles might be 
able to shed a light on other sources, such as the small $s\overline{s}$ 
component in the nucleon. 
At large angles, significant 
sensitivities to the phases of the couplings
provide insights to the small, but still sizable {\it s}- 
and {\it u}-channel resonance contributions. These results hold
also in the case of OZI suppression relevance. 
Some of the phase sets for those parameters ($a$ and $b^\prime$), offer
also a test of deviations from the exact $SU(6)\otimes O(3)$ symmetry.

Finally, we would like to point out a few possible weak points
in this work, as summarized below: 

i) The {\it s}- and {\it u}-channel contributions are 
calculated within a quark model, while the Pomeron exchange model
is based on the Regge phenomenology. So, the consistency
between the two frames remains to be investigated.
	
ii) The form factor $F_1(t)$ for the Pomeron coupling to the proton
has no relation with the exponential form factor in the quark model 
which comes naturally from the spatial integrals over the baryon 
wavefunctions for the resonance excitation and the $\pi^0$ exchange 
contributions.

Upcoming data from JLAB~\cite{JLAB} are expected to allow 
to disentangle various components of the reaction mechanism investigated 
here. Confrontation between the present theoretical predictions and
data might show the limits of the non-relativistic constituent quark
approach presented here and the need for more sophisticated and a
fully relativistic formalism.

%
%  %%  %%  %%  %%  %%  %%  %% ACKNOWLEDGMENTS
%  
\section*{ Acknowledgments} 
One of us (Q. Z.) expresses his thanks to IPN-Orsay and the 
`` Bourses de Recherche CNRS-K.C. WONG" for financial support.
Two of us (Q. Z. and B. S.) would like to express appreciations 
to Zhenping Li for his interest in this work and much encouragement. 
Q. Z. acknowledges the beneficial discussions with C. Bennhold
and W. Kloet. 
\newpage
%  %%  %%  %%  %%  %%  %%  %%  APPENDIX
%
\section*{ Appendix}
Generally, the helicity amplitudes can be explicitly written as:
\begin{eqnarray}
H_{a\lambda_V}\equiv H^r_{a\lambda_V}+i H^i_{a\lambda_V} \,
\end{eqnarray}
where $H^r_{a\lambda_V}$ and $H^i_{a\lambda_V}$ represent the 
real and imaginary part of the amplitude, respectively. 
$\lambda_V$ ($=0, \ \pm 1 $) is the helicity of the vector meson.
With the $\Gamma$ and $\omega$ matrices given by Ref.~\cite{tabakin}, 
the four single and one double polarization asymmetries of the spin observables 
can be expressed 
explicitly as follows: 

Polarized photon asymmetry
\begin{eqnarray}
\check{\Sigma}&=&\frac 12\langle H|\Gamma^4\omega^A|H \rangle \nonumber\\
&=&\frac 12\{ -H^r_{1-1}H^r_{41}-H^i_{1-1}H^i_{41}
+H^r_{10}H^r_{40}+H^i_{10}H^i_{40}\nonumber\\
&&-H^r_{11}H^r_{4-1}-H^i_{11}H^i_{4-1}
+H^r_{2-1}H^r_{31}+H^i_{2-1}H^i_{31}\nonumber\\
&&-H^r_{20}H^r_{30}-H^i_{20}H^i_{30}
+H^r_{21}H^r_{3-1}+H^i_{21}H^i_{3-1} \} \ .
\end{eqnarray}
Polarized target asymmetry
\begin{eqnarray}
\check{T}&=&-\frac 12\langle H|\Gamma^{10} \omega^1|H \rangle\nonumber\\
&=&\sum_{\lambda_V} \{ H^r_{1\lambda_V} H^i_{2\lambda_V}
-H^i_{1\lambda_V} H^r_{2\lambda_V}
+H^r_{3\lambda_V} H^i_{4\lambda_V}
-H^i_{3\lambda_V} H^r_{4\lambda_V} \}\ .
\end{eqnarray}
Polarized vector meson asymmetry
\begin{eqnarray}
\check{P}_V&=&\frac 12 \langle H|\Gamma^1\omega^3|H\rangle \nonumber\\
&=& \frac{\sqrt{3}}2 \{H^r_{1-1}H^i_{10}-H^i_{1-1}H^r_{10}
+H^r_{2-1}H^i_{20}-H^i_{2-1}H^r_{20}\nonumber\\
&& H^r_{3-1}H^i_{30}-H^i_{3-1}H^r_{30}+H^r_{4-1}H^i_{40}-H^i_{4-1}H^r_{40}\} \ .
\end{eqnarray}
Recoil polarization asymmetry
\begin{eqnarray}
\check{P}_{N^\prime}&=&\frac 12\langle H|\Gamma^{12}\omega^1|H\rangle\nonumber\\
&=&\sum_{\lambda_V} \{H^i_{3\lambda_V}H^r_{1\lambda_V}
-H^r_{3\lambda_V}H^i_{1\lambda_V}
+H^i_{4\lambda_V}H^r_{2\lambda_V}
-H^r_{4\lambda_V}H^i_{2\lambda_V} \} \ .
\end{eqnarray}

The explicit expression for the component ${\mathcal C}^{\gamma N}_{zz}$ 
in the beam-target double polarization asymmetry is given by:
\begin{eqnarray}
{\mathcal C}^{\gamma N}_{zz}&=
&\frac 12\langle H|\Gamma^9\omega^1|H \rangle\nonumber\\
&=& \frac 12\sum_{\lambda_V} 
\{ H^r_{1\lambda_V}H^r_{1\lambda_V}+H^i_{1\lambda_V}H^i_{1\lambda_V}
-H^r_{2\lambda_V}H^r_{2\lambda_V}-H^i_{2\lambda_V}H^i_{2\lambda_V}\nonumber\\
&&+H^r_{3\lambda_V}H^r_{3\lambda_V}+H^i_{3\lambda_V}H^i_{3\lambda_V}
-H^r_{4\lambda_V}H^r_{4\lambda_V}-H^i_{4\lambda_V}H^i_{4\lambda_V} \} \ .
\end{eqnarray}

It is worth noting that the helicity amplitudes can be explicitly 
related to the density matrices which can be measured 
in experiments~\cite{schilling}. 
\newpage
%
%  %%  %%  %%  %%  %%  %%  %%  REFS
%

%
%%%%%%%%%%%%%%%%%%%%%%%%%%%%%%%
% 
\end{document}